\documentstyle[12pt]{article}

\newcommand{\Abar}{\not{\!{\!A}}}

\newcommand{\pabar}{\not{\!\partial}}
\newcommand{\Od}{{\cal O}}

\newcommand{\Tr}{\mbox{Tr}}
\newcommand{\im}{\mbox{Im}}

\newcommand{\Dbar}{\not{\!{\!D}}}

\newcommand{\ZP}[1]{{\em Z.\ Phys.\ }{\bf #1}}
\def\IN{\relax{\rm I\kern-.18em N}}
\def\IR{\relax{\rm I\kern-.18em R}}
\font\cmss=cmss12 \font\cmsss=cmss12 at 7pt
\def\IZ{\relax\ifmmode\mathchoice
{\hbox{\cmss Z\kern-.4em Z}}{\hbox{\cmss Z\kern-.4em Z}}
{\lower.9pt\hbox{\cmsss Z\kern-.4em Z}}
{\lower1.2pt\hbox{\cmsss Z\kern-.4em Z}}\else{\cmss Z\kern-.4em Z}\fi}

\def\inbar{\,\vrule height1.5ex width.4pt depth0pt}
\def\IC{\relax\hbox{$\inbar\kern-.3em{\rm C}$}}

\def\ddot{\mathaccent"707F}

\begin{document}
\input epsf
\title{Particle production from nonlocal gravitational effective action}
\author{Antonio Dobado \\
Departamento de F\'{\i}sica Te\'orica \\
Universidad Complutense de Madrid\\
 28040 Madrid, Spain\\ and \\
Antonio L. Maroto\\
Astronomy Centre \\ 
University of Sussex,\\
Brighton, BN1 9QJ, U.K.}

\date{\today}

\maketitle
\begin{abstract}
In this paper we show how the nonlocal effective action
 for gravity, 
obtained after integrating out the matter fields, can be used to compute
 particle
production  and spectra for different space-time metrics. Applying
this technique to several examples, we find that
the  perturbative calculation of the effective action up to second order 
in curvatures yields exactly the same results for the total number of 
particles as the Bogolyubov transformations method, in the case of masless
scalar fields propagating in a Robertson-Walker space-time. 
Using an adiabatic approximation we also obtain the corresponding spectra
and compare the results with the traditional 
WKB approximation.    
 
\end{abstract}
\newpage
%\baselineskip 0.83 true cm
%\textheight 20 true cm
%\newpage

\section{Introduction}

In recent years the phenomenon of particle creation from classical sources
has experienced a growing interest, mainly motivated by its numerous
applications in cosmology, but also in other areas of physics. In cosmology, 
it plays a fundamental role in the mechanism of reheating after inflation
\cite{Linde}
which is believed to be responsible for the creation of almost all the
particles that populate the universe today. In the reheating models, the 
oscillations of an homogeneous scalar field (inflaton) around the minimum of
its potential give rise to an explosive creation of  
a large amount of particles. On the other hand, the same methods
are applied to the generation of primordial density inhomogeneities
in the early universe that
later on grew to create the present galactic structure \cite{brand}. 
In addition, 
the cosmological expansion can give rise to the production of an 
stochastic background of gravitational waves \cite{graviton}. 
Bounds on the density of these
waves are very useful to constraint the different cosmological models
\cite{Barrow}. In all these
applications,  the method which is used for the calculation of the 
rates and spectra of the particles produced is the traditional 
mode-mixing Bogolyubov technique \cite{Birrell}.  

On the other hand the  notion of effective action (EA) has proved to be 
a very useful 
tool for the development of the so called phenomenological lagrangians. 
Typically, effective 
actions are obtained in theories with heavy and light fields by 
functional integration 
of the heavy modes to find the effective low-energy theory for 
the light modes after 
some momentum expansion. Usual applications of those techniques 
include low-energy hadron 
dynamics (the so called Chiral Perturbation Theory), the 
symmetry breaking sector of the 
standard model, and low-energy quantum gravity 
(see \cite{libro} for a recent review and 
references therein). Effective actions use to have a real and, in general
divergent part, that give rise to modifications of the classical equations
of motion due to quantum effects. Eventually, the  corresponding 
vacuum  solutions 
could not exhibit  some of the  symmetries of the classical theory, thus 
giving rise to the well-known 
phenomenon of spontaneous symmetry breaking. 
In addition, nonlocal finite 
terms also appear in the EA which contribute to the imaginary part. 
This imaginary part is physically  
important since it is connected with the possibility of having 
particle production \cite{Hu,Schwinger}. By this we 
mean the production of the quanta corresponding to the 
fields that have been integrated out. 

In this paper, we consider the production of scalar  particles from classical
gravitational backgrounds from the effective action point of view.
We show how a perturbative calculation 
up to second order in the
curvatures in the case of masless scalar fields, reproduces the 
well-known general results of particle production in Robertson-Walker
space-times and  can give rise to the
exact amount of particles at least in the models we have considered.
The paper is organized as follows: in Section 2,  we review 
the Euler-Heisenberg
lagrangian for QED, but paying special attention to its nonlocal part.
We show how the perturbative calculation up to second order
in the coupling constant yields
the correct expression for the imaginary part in the masless case.
In section 3, we introduce the nonlocal gravitational effective action
for scalar fields and discuss some of the conditions for its application.
Section 4 is devoted to the actual calculation of the total number
of particles produced due to the expansion in several Robertson-Walker
models and the results are compared with those
obtained by the Bogolyubov technique. 
In Section 5  we study how to obtain the spectrum of the particles
and compare the results with the WKB approximation. Finally, Section 6
contains the main conclusions of the work.

\section{The nonlocal Euler-Heisenberg lagrangian}
Let us consider the well-know Euler-Heisenberg lagrangian for
QED  in flat space-time \cite{EulerHeisenberg}. When the 
momentum $p$ of photons  is
much smaller than the electron mass $M$, the one-loop effects, such as 
vacuum
polarization, can be taken into account by adding local non-linear terms 
to the classical
electromagnetic lagrangian. Consider the QED effective action given by:
\begin{eqnarray}
e^{iW[A]}&=&\int [d\psi][d\overline\psi]e^{-\frac{i}{4}\int d^4x 
F_{\mu\nu}F^{\mu\nu}}
\exp\left(i\int d^4x\overline \psi (i\Dbar-M+i\epsilon)\psi\right)\nonumber \\
&=&
e^{ -\frac{i}{4}\int d^4x F_{\mu\nu}F^{\mu\nu}} \det (i\Dbar-M+i\epsilon)
\label{qedea}
\end{eqnarray}
where as usual
 $\Dbar=\gamma^\mu(\partial_\mu-ieA_\mu)$. From (\ref{qedea}) we can write
the effective action as:
\begin{eqnarray}
W[A]=-\frac{1}{4}\int d^4x F_{\mu\nu}F^{\mu\nu}-i\Tr \log((i\Dbar-M
+i\epsilon))
\end{eqnarray}
Expanding in a formal way the logarithm we obtain:
\begin{eqnarray}
W[A]=-\frac{1}{4}\int d^4x F_{\mu\nu}F^{\mu\nu}
+i\sum_{k=1}^{\infty} \frac {(-e)^k}{k}\Tr[(i{{\pabar}} -M)^{-1} \Abar]^k 
\end{eqnarray}
Using dimensional
regularization, it is possible to find the following 
expression up to quadratic
terms in the photon field:
\begin{eqnarray}
W[A]&=&
\int d^4x \left[-\frac{1}{4}F^{\mu\nu}F_{\mu\nu}-\frac{e^2}{3(4\pi)^2}\Delta 
F^{\mu\nu}F_{\mu\nu}
-\frac{2e^2}{(4\pi)^2}F^{\mu\nu}\left(-\frac{2}{3}\frac{M^2}{\Box}\right. 
\right.\nonumber \\
&-&\left.\left. \frac{1}{6}\left(1-2\frac{M^2}{\Box}\right)F(-\Box;M^2)\right) 
F_{\mu\nu}\right]
+\Od(A^4)
\label{ehea}
\end{eqnarray}
where $\Delta=N_\epsilon-\log (M^2/\mu^2)$,
$N_{\epsilon}=2/\epsilon-\gamma +\log 4\pi$
 is the well known constant appearing in 
dimensional regularization and we have used the 
expression:
\begin{eqnarray}
F(-\Box;M^2)F_{\mu\nu}(x)=\int d^4y\frac{d^4p}{(2\pi)^4}e^{ip(x-y)}F(p^2;M^2)
F_{\mu\nu}(y)
\end{eqnarray}
with:
\begin{eqnarray}
F(p^2;M^2)=2+\int_0^1 dt \log\left(1-\frac{p^2}{M^2}t(1-t)\right)
\end{eqnarray} 
In the $p^2>4M^2$ case, this function can be written as:
\begin{eqnarray}
F(p^2;M^2)=\sqrt{1-\frac{4M^2}{p^2}}\log \frac{\sqrt{1-\frac{4M^2}{p^2}}+1}
{\sqrt{1-\frac{4M^2}{p^2}}-1}
\label{mandelstam}
\end{eqnarray}
In a similar way, the inverse operator $1/\Box$ can be defined with the usual 
boundary 
conditions on the fields as:
\begin{eqnarray}
\frac{1}{-\Box}F_{\mu\nu}(x)=\int d^4y\frac{d^4p}{(2\pi)^4}e^{ip(x-y)}
\frac{1}{p^2+i\epsilon}F_{\mu\nu}(y)
\end{eqnarray}
The expression (\ref{ehea}) for the EA has a regular massless limit. In fact, 
for 
small $p$ compared with $M$, the Mandelstam function $F(p^2;M^2)$ behaves as:
\begin{eqnarray}
F(p^2;M^2)= -\log \left(\frac{M^2}{-p^2-i\epsilon}\right)+\Od(M^2)
\label{mandaprox}
\end{eqnarray}
From (\ref{ehea}) we can see that the only contributions in the massless limit 
are
those coming, on one hand from the  $\Delta$ factor and, on the other hand, 
from the 
Mandelstam function. Both logarithmic contributions equal, up to  sign, 
so that they cancel each other and we obtain:
\begin{eqnarray}
W[A]=\int d^4x \left(-\frac{1}{4}F^{\mu\nu}F_{\mu\nu}
-\frac{e^2}{3(4\pi)^2} F^{\mu\nu}\Gamma(\Box)F_{\mu\nu}\right)+\Od(A^4)
\label{eheam0}
\end{eqnarray}
where we have used the following notation:
\begin{eqnarray}
\Gamma(\Box)=N_\epsilon-\log\left(\frac{\Box}{\mu^2}\right)
\label{gama}
\end{eqnarray}
to be understood as in the previous cases through the corresponding Fourier 
transform, with
the $i\epsilon$ factor as shown in (\ref{mandaprox}). We see that the
masless limit of the EA is a nonlocal but analytical functional
in the gauge curvatures $F_{\mu\nu}$.
 
The EA (\ref{ehea}) allows us to derive in an exact fashion the photon 
two-point  
one loop Green functions. This, in turn, 
allows us to obtain for example the vacuum polarization. In the massive 
case, the EA can be 
expanded
as a power series in $p^2/M^2$, and also in $A$ to obtain the well-known 
local Euler-Heisenberg
lagrangian 
\cite{EulerHeisenberg}:
\begin{eqnarray}
{\cal L}_{eff}&=&-\frac{1}{4}F^{\mu\nu}F_{\mu\nu}-\frac{e^2}{3(4\pi)^2}
\Delta F^{\mu\nu}F_{\mu\nu}
-\frac{e^2}{15(4\pi)^2 M^2}F^{\mu\nu}\Box F_{\mu\nu}\nonumber \\
&+&\frac{e^4}{90(4\pi)^2M^4}\left((F^{\mu\nu}F_{\mu\nu})^2+
\frac{7}{4}(F^{\mu\nu}\tilde F_{\mu\nu})^2\right) +\Od\left(\frac{p^2}
{M^2}\right)^3
+\Od(A^6)
\label{ehlag}
\end{eqnarray} 

The EA (\ref{ehea}) possesses a non-vanishing imaginary part
coming from the Mandelstam function (\ref{mandelstam}). This
imaginary part provides the pair production rate \cite{Schwinger}.  
In the massless case  (\ref{eheam0}) we get:
\begin{eqnarray}
\im \;W[A]=\im \int d^4x {\cal L}_{eff}=
-\frac{e^2}{48\pi}\int d^4x \; d^4y\frac{d^4p}{(2\pi)^4}
e^{ip(x-y)}F_{\mu\nu}(x)F^{\mu\nu}(y)\theta(p^2)
\end{eqnarray}
where:
\begin{eqnarray}
\theta(x)=\left \{ \begin{array}{cc}
0 & x<0 \\
1/2 & x=0 \\
1 & x>0 
\end{array}\right.
\end{eqnarray}
The $1/2$ value arises as a consequence of the $-i\epsilon$ factor in
 (\ref{mandaprox}).
For constant electric fields and in absence of magnetic fields, the 
previous expression gives
the probability per unit time and unit volume that at least one 
electron-positron pair
is created by the electric field:
\begin{eqnarray}
p\simeq 2 \im\; {\cal L}_{eff}=\frac{e^2}{24\pi}\vec E^2
\label{pert}
\end{eqnarray}
Let us compare this result with exact expression for the imaginary part
obtained by Schwinger \cite{Schwinger}:
\begin{eqnarray}
p\simeq 2 \im\; {\cal L}_{eff}
=\frac{e^2E^2}{4\pi^3}\sum_{n=1}^\infty \frac{1}{n^2}e^{-\frac{m^2n\pi}{eE}}
\label{eres}
\end{eqnarray}
The dependence in the electric fields appears in both a quadratic term and an
non-analytical contribution $\exp(-m^2n\pi/eE)$. This latter term shows the 
importance of the non-perturbative effects in the particle production
phenomenon \cite{Linde}.
 However, in the masless limit
the non-analytical pieces disappear and the result exactly agrees with
the perturbative calculation in (\ref{pert}). Notice that in this case, gauge
invariance and the dimension of the effective lagrangian constraint the result 
to be quadratic in $eE$ and that is the reason why the second order 
perturbative calculation gives rise to the exact result. Accordingly, in
the masless limit, the perturbative calculation can provide,
in some cases, all the relevant information about the particle production
processes.
 In the gravitational
case that we will study in the  next sections, we will show that the same 
effect takes place.

\section{The  effective action for gravity}
Let us consider a real scalar field in a curved space-time
with an arbitrary non-minimal coupling to the curvature.
The 
corresponding classical action is given by:
\begin{eqnarray}
S[\phi]=-\frac{1}{2}\int d^4x\sqrt{g} \phi\left( \Box+m^2+\xi R \right) \phi
\label{acsc}
\end{eqnarray}
where:
\begin{eqnarray}
\Box \phi=g^{\mu\nu}\nabla_{\mu}\partial_{\nu}\phi
=\frac{1}{\sqrt{g}}\partial_{\mu}\left(g^{\mu\nu}\sqrt{g}
\partial_{\nu}\phi\right)
\end{eqnarray}
The EA for the gravitational fields that arises after integrating out the  
real scalar matter fields is given by the following expression in Lorentzian 
signature:
\begin{eqnarray}
\langle 0,out \vert 0,in \rangle &=& 
Z[g_{\mu\nu}]=e^{iW[g_{\mu\nu}]}
= \int [d\phi]e^{iS[g_{\mu\nu},\phi]}\nonumber \\
&=&
 \int [d\phi]e^{-\frac{i}{2}\int dx\sqrt{g}
\phi(\Box+m^2+\xi R -i\epsilon)\phi}=
(\det O)^{-\frac{1}{2}}
\label{eaes}
\end{eqnarray}
where  
$O_{xy}(m^2)=(-\Box_y-m^2-\xi R(y) +i\epsilon)\delta^0(x,y)$ 
with $\delta^0(x,y)$ being  
the covariant delta $\delta^0(x,y)=g^{-1/2}(x)\delta(x,y)$. Thus we see
that, following the analogy with flat space-time we could interpret
$Z[g_{\mu\nu}]$ as the vacuum persistence amplitude.
 Thus we have 
\begin{eqnarray}
W[g_{\mu\nu}]=\frac{i}{2}\log \; \det O(m^2)=\frac{i}{2}\Tr \;\log O(m^2)
\label{esceff}
\end{eqnarray}
In this expression we have integrated the scalars but the gravitational field
is treated classically. Accordingly, this EA must be added to the classical 
action for 
the gravitational field and it includes
the quantum effects due to the matter fields. In addition, (\ref{eaes}) 
is the generating functional of the Green functions containing  scalar 
loops only and
external gravitational legs. 

Once one knows the EA at least in some limit, we have all
the information concerning the semi-classical
gravitational evolution in this limit. As we mentioned in the
introduction,  
the EA could have
a non-vanishing imaginary part, which is related to the pair 
production probability. 
In fact, the probability $P$ that at least one pair particle-antiparticle
is created by the gravitational field is given by \cite{Schwinger}:
\begin{eqnarray}
P=1- \vert\langle 0,out \vert 0,in \rangle\vert ^2_{g_{\mu\nu}} =
1-\vert e^{iW[g_{\mu\nu}]}\vert^2=1-e^{-2 Im \; W[g_{\mu\nu}]}
\end{eqnarray}
for small values of $W[g_{\mu\nu}]$ we have:
\begin{eqnarray}
P\simeq 2\im \; W[g_{\mu\nu}]
\label{improd}
\end{eqnarray}

Concerning the applicability of this equation, let us compare the
EA method with the traditional Bogolyubov technique. 
The classical equations
of motion for the scalar field are:
\begin{eqnarray}
(\Box +m^2+\xi R)\phi=0
\end{eqnarray}
Unlike flat space-time, there is no natural set of mode solutions to this
equation, rather we can expand its solutions in different ways, i.e:
\begin{eqnarray}
\phi=\sum_k(a_k u_k +a^\dagger_k u_k^*)=\sum_k(\bar a_k \bar u_k 
+\bar a^\dagger_k \bar u_k^*)
\end{eqnarray}
Each of these expansions will give rise to different Fock spaces when
interpreting the coefficients $a_k, a^\dagger_k$ and $\bar a_k, 
\bar a^\dagger_k$ as creation and
annhilation operators. A problem arises when we try to identify
which of these Fock spaces corresponds to our usual notion of particle.
In general, this question can only be answered when we have a high degree
of symmetry (conformal invariance) or if the space-time is flat in the 
asymptotic {\it in} and {\it out} regions. However in most of the interesting 
situations, these two conditions are not present. A solution to this
problem was suggested in a series of works (see \cite{Birrell}
and references therein) in which the notion
of adiabatic vacuum is introduced. In the cosmological space-times 
in which we will be mainly interested, in order to define an adiabatic
vacuum it is only required that asymptotically in the past and
in the future the rate of expansion vanishes, i.e. $\dot a/a\rightarrow 0$
with $a(t)$ the universe scale factor. Expressing this statement in
a covariant way, it would be equivalent to require
that the curvatures and all their covariant derivatives vanish in the 
far past and future.

In the effective action approach, $Z[g_{\mu\nu}]$ can be interpreted
as vacuum persistence amplitude in principle only  when
the vacuum states $\vert 0,in\rangle$ and $\vert 0, out\rangle$ can be
defined in regions with a temporal separation \cite{parker0}. 
When this does not
occur, it is not obvious what is the interpretation of the effective action.
However, we will show in the following, that the naive calculation of the
effective action, in those situations in which an adiabatic vaccum can
be defined although  the space-time is not asymptotically  Minkowskian,
yield the same result for the particle production as the standard Bogolyubov
technique. As a consequence, in these cases, we could try to
interpret $Z[g_{\mu\nu}]$ as adiabatic vacuum persistence amplitude.

The nonlocal effective action for gravity has been evaluated in 
different works using several techniques. Thus in \cite{Duff} it
was suggested what would be the form of the two-point form factors. In 
\cite{vilko} the effective action is derived by means of the so called
covariant perturbation theory, valid in asymptotically flat manifolds,
in \cite{avra} the same result is obtained by means of the partial
resummation of the Schwinger-DeWitt series. The result in all these
cases up to second order in curvatures can be written in the masless case as:
\begin{eqnarray}
W[g_{\mu\nu}]&=&\frac{1}{32\pi^2}\int d^4 x \sqrt{g}
\left(\frac{1}{180}R^{\mu\nu\lambda\rho}(x)\Gamma(\Box)
R_{\mu\nu\lambda\rho}(x)-\frac{1}{180}R^{\mu\nu}(x)\Gamma(\Box)R_{\mu\nu}(x)
\right.  \nonumber\\
&+& \left. 
\frac{1}{2}\left(\frac{1}{6}-\xi\right)^2 R(x)\Gamma(\Box)R(x)\right)
+\Od ({\cal R}^3)
\label{nolocal}
\end{eqnarray}  
where the form factor $\Gamma(\Box)$ is given in (\ref{gama}).
The local finite pieces as usual depend on the different renormalization
squemes and they are not relevant for our calculations, although in
general their coefficients are important to fix the form of the linear
terms in the trace anomaly.  
The nonlocal contributions are in any case unambiguous.
An appropriate representation of the nonlocal form factors is 
provided by  the use of the Riemann normal coordinates (the details of this
approach 
will be given elsewhere \cite{DoMa}).
Thus, taking normal coordinates  ($x^\mu$) with origin at $y_0$
the action of the form factors is understood through the correspondig
Fourier transform:
\begin{eqnarray}
{\cal R}(y_0)\; \log\left(\frac{\Box}{\mu^2}\right)\;{\cal R}(y_0)=
\int d^4x \frac{d^4p}{(2\pi)^4}e^{ip x}{\cal R}(y_0)\log 
\left(\frac{-p^2-i\epsilon}{\mu^2}\right){\cal R}(x)
\label{act}
\end{eqnarray}
and ${\cal R}$ denotes generically the scalar curvature, the Ricci or 
Riemann tensors. 

For the sake of simplicity we will study massless scalar particles 
propagating in a
cosmological background, whose metric is that of 
Friedmann-Robertson-Walker (FRW): 
\begin{eqnarray}
ds^2=dt^2-a(t)^2\left(\frac{dr^2}{1-Kr^2}+r^2d\theta^2+r^2\sin^2 
(\theta)d\phi^2\right)
\label{rw}
\end{eqnarray}
where $K$ determines the spatial curvature sign \cite{Wei}  and $a(t)$
is the universe scale factor.   

The EA imaginary part comes from the logarithms in (\ref{nolocal}). 
Due to the homogeneity and isotropy of space in the present case, 
the different curvatures
appearing in that expression only depend on the time coordinate. Thus, we can 
perform the spatial coordinates integration in (\ref{act}) and 
generically we will obtain:
\begin{eqnarray}
&\im & \int d^4x \frac{d^4p}{(2\pi)^4}e^{ip x}{\cal R}(y_0)\log 
\left(\frac{-p^2-i\epsilon}{\mu^2}\right){\cal R}(x)\nonumber\\
&=&
\im \int dx^0 \frac{dp_0}{(2\pi)}e^{ip_0 x^0}
{\cal R}(y_0)\log\left(\frac{-p_0^2-i\epsilon}{\mu^2}\right){\cal R}(x^0)
\nonumber \\
&=&-\pi{\cal R}(y_0){\cal R}(y_0)
\label{frwim}
\end{eqnarray}
 Let us momentarily 
consider a general metric, not necessarily FRW. It is easy to see from
the first term in this equation that when the metric is static, 
i.e only depending on
spatial coordinates, the argument in the logarithm would only contain 
$\vec p^2-i\epsilon$.
Therefore the imaginary part would be zero and we would recover 
the well-known result of absence
of particle production in general (inhomogeneous) static backgrounds.

\begin{itemize}
\item {\it FRW metrics}

Returning to the FRW metric
we obtain from (\ref{frwim}) the general expression:
\begin{eqnarray}
\im \; W[g_{\mu\nu}]&=&\frac{1}{32\pi}\int d^4x \sqrt{g} 
\left(\frac{1}{180}R^{\mu\nu\lambda\rho}R_{\mu\nu\lambda\rho}-
\frac{1}{180}R^{\mu\nu}R_{\mu\nu}\right.\nonumber\\
&+&
\left.\frac{1}{2}\left(\frac{1}{6}-\xi\right)^2 R^2\right)
+\Od ({\cal R}^3)
\label{imea}
\end{eqnarray}
This result is only valid for homogeneous and isotropic metrics.
Comparing this result with the divergences, we see
that both have the same form. 
Notice that (\ref{imea})
is a linear combination of $R_{\mu\nu\lambda\rho}^2$, $R_{\mu\nu}^2$ and $R^2$,
but we can choose a different basis to write it. In particular, 
we can take the one 
made out of $R^2$, 
$C_{\mu\nu\lambda\rho}^2$ and $E$, where $C_{\mu\nu\lambda\rho}$ is 
the Weyl tensor and
$E=R_{\mu\nu\lambda\rho}^2-4R_{\mu\nu}^2+R^2$ is the Gauss-Bonnet term.
In this basis we have:
\begin{eqnarray}
a_1R_{\mu\nu\lambda\rho}^2+a_2R_{\mu\nu}^2+a_3R^2&=&-\left(a_1
+\frac{1}{2}a_2\right)E+\left(2a_1+\frac{a_2}{2}\right)C^2\nonumber \\
&+&\left(\frac{1}{3}a_1+\frac{1}{3}a_2+
a_3\right)R^2 
\label{curvaturas}
\end{eqnarray}
In our case, $a_1=-a_2$.
On the other hand, the FRW metric is locally conformal to the Minkowski metric
and hence its Weyl tensor  vanishes. Therefore (\ref{imea})
only contains the scalar curvature and  the Gauss-Bonnet terms, 
but the integral of the latter also vanishes in the class of 
asymptotically flat metrics. Moreover, the asymptotic flatness
is not a neccessary condition for the Gauss-Bonnet term
to vanish and, in fact, examples can be found which are not
asymptotically flat, but still they have a zero Gauss-Bonnet
term contribution (see below). To summarize, the
imaginary part in these cases reduces to:
\begin{eqnarray}
\im \; W[g_{\mu\nu}]&=&\frac{1}{32\pi}\int d^4x \sqrt{g} 
\frac{1}{2}\left(\frac{1}{6}-\xi\right)^2 R^2
+\Od ({\cal R}^3)
\label{frwims}
\end{eqnarray} 

\item {\it Conformal coupling}

In the conformal case ($\xi=1/6$) it is evident from the above expression that
the EA imaginary part
is zero and accordingly there will be no particle production. This is 
a well-known result and has been proved by studying the positive-energy 
modes of the 
corresponding Klein-Gordon equation \cite{Parker1} for the scalar field. 
The EA provides 
in this case a simple
way to prove a general result.

\item {\it Radiation dominated universe}

But conformal invariance is not the only case in which there is no 
particle production
in a FRW background. From the above arguments, we have seen that the 
only piece contributing
to the EA imaginary part is the $R^2$ term. If this term vanishes, 
there would not be 
particle creation. For a FRW metric with $K=0$ this implies the 
following condition:
\begin{eqnarray}
\dot H=-2H^2
\end{eqnarray}
where $H=\dot a/a$ is the Hubble parameter. The solution is simply: 
$a(t)=A(t-t_0)^{1/2}$ with $A$ and $t_0$ arbitrary constants. In fact
taking traces in the Einstein equations (with the stress tensor 
corresponding to a
perfect fluid), it is obvious that $R=0$ implies $R=8\pi G (3p-\rho)=0$,
with $p$ and $\rho$ the pressure and density of the fluid. Accordingly
$\rho=3p$, which is nothing but the state equation for a fluid of highly 
relativistic
particles. Therefore a radiation dominated universe is a stable solution 
of Einstein equations
against pair emission. This result was 
obtained in \cite{Parker1,Parkernew,Parker2}  by
means of the Bogolyubov technique, where
in order to circumvent the problem of the initial singularity,
it was assumed that when $t\rightarrow t_P$ with $t_P$ the Planck time, 
the scale factor smoothly tends to a constant. 
This allows us to define
an initial vacuum state in the problem. Again the {\it out} vacuum is
chosen as  an adiabatic vacuum. 
Notice that in this case the Gauss-Bonnet term can also be neglected.

\item {\it Homogeneous anisotropic metrics}

Consider now a general homogeneous but anisotropic metric of the Bianchi 
type I:
\begin{eqnarray}
ds^2=C^2(\eta)(d\eta^2-g_{ij}(\eta)dx^idx^j)
\end{eqnarray}
where the 3-metric $g_{ij}$ only depends on the time coordinate. Since, 
as it happened with
the FRW, the curvatures  only depend on the time coordinate, it 
is possible to explicitly 
perform the spatial coordinate integration in (\ref{act}). Therefore we
obtain the same combination of curvature tensors as in (\ref{imea}) 
for the EA imaginary part. 
In this case
the metric is not conformal to the Minkowski one and accordingly 
it is not possible to
drop the  Weyl term from (\ref{curvaturas}). The Gauss-Bonnet 
term continues vanishing under the same assumptions about the metric. 
To summarize, the resulting EA imaginary part can be written 
for this kind of metrics as:
\begin{eqnarray}
\im \; W[g_{\mu\nu}]&=&\frac{1}{32\pi}\int d^4x \sqrt{g}\left(
\frac{1}{120}C_{\mu\nu\rho\sigma}
C^{\mu\nu\rho\sigma}+
\frac{1}{2}\left(\frac{1}{6}-\xi\right)^2 R^2\right)
\nonumber \\&+&\Od ({\cal R}^3)
\label{imbi}
\end{eqnarray}

This result agrees with that of Zel'dovich and Starobinski \cite{Ze} (see also
\cite{Bi}) obtained by 
using standard Bogolyubov techniques. In fact,  assuming 
$g_{ij}=\delta_{ij}(1+h_i(\eta))$, 
neglecting terms of order $\Od(h^3)$ in (\ref{imbi}) and imposing 
that asymptotically the anisotropies vanish, we recover 
their results.   

\end{itemize}

From the above expressions we can extract another consequence. Particle 
production only takes
place when curvature is non-vanishing, i.e, in the presence of a genuine 
gravitational
field and not merely by means of a coordinate change as it happens for an 
accelerated
observer \cite{Birrell}, in  the latter case the creation could be 
considered as
fictitious. Therefore, for the boundary conditions in the space-time geometry 
that we 
mentioned before, the EA provides an invariant criterium (independent of the
observer) to decide when particle production takes place.

\section{Specific examples with minimal coupling}

In order to illustrate the previous results we will show several examples in 
which the 
EA allows us to make physical predictions. In some cases it will be possible to
compare these results with those obtained by means of the traditional 
Bogolyubov 
transformations. Exact results from the Bogolyubov
transformation have been obtained for a very limited number
of models in the literature.
\begin{itemize}
\item{\it Model 1}
\end{itemize}

We will now consider a  {\it complex} scalar field and the FRW  metric
with $K=0$.
It will be useful, in order to compare with other 
results, to work with the new time coordinate defined by:  
\begin{eqnarray}
\tau=\int^t a^{-3}(t')dt'
\label{tau}
\end{eqnarray}
that allows us to write the d'Alembertian operator acting on time dependent 
functions as:
\begin{eqnarray}
\Box f(\tau)=\frac{1}{a^6}\partial_\tau\partial_\tau f(\tau)
\label{dal}
\end{eqnarray}

First we consider the model proposed in \cite{Parker2}. The scale
factor is given by:
\begin{eqnarray}
a^4(\tau)\simeq a_1^4+e^{\tau/s}\left((a_2^4-a_1^4)
(e^{\tau/s}+1)+b\right)(e^{\tau/s}+1)^{-2}
\label{park}
\end{eqnarray}
For $\tau \rightarrow \infty(-\infty)$, $a(\tau)$ smoothly tends to a 
constant $a_2(a_1)$, i.e,
it is possible to unambiguously define initial and final vacuum states. 
On the other 
hand, 
$a_1,a_2, s$ and $b$ are arbitrary parameters. For $a_2>>a_1$ and using 
quantum mechanics 
methods, it is possible to calculate the Bogolyubov coefficients and hence the
number density of produced particles \cite{Parker2}:
\begin{eqnarray}
<N_k>=\frac{1}{e^{4\pi s a_1^2 k}-1}
\label{bogpar}
\end{eqnarray}
where  the number  of created pairs per unit
coordinate volume and unit trimomentum volume in the $k$ mode is related to
the Bogolyubov coefficients by means of: \\
$<N_k>=\vert \beta_k \vert^2$ \cite{Birrell}

The relation between $<N_k>$ and the pair production probability 
 per unit coordinate volume $p_{BOG}$
is given by this expression \cite{Damour}:
\begin{eqnarray}
p_{BOG}\simeq \int \frac{d^3 k}{(2\pi)^3}\left[ \pm \log (1\pm <N_k>)\right]
\label{imbo}
\end{eqnarray}
where the $+$ sign is used for bosons and $-$ for fermions.

In  the present model $<N_k>$ does not depend on $a_2$ nor $b$. 
Using (\ref{imbo}) 
we find for the probability density:
\begin{eqnarray}
p_{BOG}=\int \frac{d^3 k}{(2\pi)^3} \log (1+ <N_k>)=
\int dk\frac{k^2}{2\pi^2} \log \left(1+\frac{1}{e^{4\pi s a_1^2 k}-1}\right)
\label{wbog}
\end{eqnarray}  
On the other hand, the EA method provides from (\ref{frwims}):
\begin{eqnarray}
p_{EA}\simeq 2 \frac{1}{36}\frac{1}{32\pi}\int d\tau a^6(\tau)R^2(\tau)
\label{earw}
\end{eqnarray} 
with:
\begin{eqnarray}
R(\tau)=-12\frac{\dot a^2}{a^8}+6\frac{\ddot a}{a^7}
\label{escrw}
\end{eqnarray}

It is possible to perform the integrals in (\ref{wbog}) and (\ref{earw}) in
an explicit way, so that we can compare both results at the analytical
level. They yield {\it exactly} the same result:
\begin{eqnarray}
p_{EA}=p_{BOG}=\frac{1}{5760\pi a_1^6s^3}
\end{eqnarray}
where we have taken the limits $b=0$ and $a_2\rightarrow \infty$.

\begin{itemize}
\item{\it Model 2}
\end{itemize}

The second model we will study is that proposed in \cite{Schafer}.
The scale factor is now given by:
\begin{eqnarray}
a^4(\tau)=A^2\tau^2+B^2
\end{eqnarray}
where $A$ and $B$ are arbitrary constants. In this case, space-time
is not asymptotically flat and therefore the Bogolyubov calculation
is based on the definition of adiabatic vacua.
However, the Gauss-Bonnet contribution vanishes and thus we can use again (\ref{frwims}).
The number of created pairs in the $k$ mode is given by:
\begin{eqnarray}
<N_k>=e^{-\frac{\pi B^2 k}{A}}
\end{eqnarray}
Once again both methods yield the same results for 
the probability densities:
\begin{eqnarray}
p_{BOG}=p_{EA}=\frac{7 A^3}{360 B^6\pi}
\end{eqnarray}
Since in (\ref{earw}) we have neglected higher order terms in curvatures, 
we can conclude
that in these two cases they do not contribute to the EA imaginary part.
As we found in the QED case, here again the second order 
perturbative calculation
is exact.
To check this fact,  we should calculate the complete expression for the EA
as we did  in Section 2, 
however the very same arguments used in that section suggest that 
in the absence of a
mass term, since in
both cases there is just one dimensional parameter, it is
not possible to build any other term with the appropriate
dimension.

\section{Spectrum and WKB approximation}

The traditional Bogolyubov method for particle production 
gives information, not only
on the total number of created particles, 
but also on their energy distribution. However, only in very
specific cases, closed analytical expressions can be written. 
As we have seen, the EA method provides a closed expression for the total
number of particle that is obtained from the curvatures and,
therefore, can be evaluated for arbitrary scale factors in a very easy way.
In this respect the EA method is obviously more advantageous than the 
Bogolyubov method.
However, it is not obvious how to derive the 
spectra in this formalism.
 Let us try to clarify this issue with a simple example and
compare our result with the one obtained from the traditional method.

Consider the Klein-Gordon equation for a minimally coupled massless 
complex scalar field
\begin{eqnarray}
\Box \phi=0
\end{eqnarray}
Introducing the FRW metric  (\ref{rw})
with time coordinate $\tau$ and $K=0$, we look 
for solutions by means of variable separation
$\phi(\tau,\vec x)=\chi_k(\tau)e^{i\vec k \vec x}$.
Hence the temporal equation can be written as:
\begin{eqnarray}
\frac{d^2 \chi_k}{d\tau^2}+a^4(\tau)k^2\chi_k=0
\label{tkg}
\end{eqnarray}
where $k^2=\vec k \vec k$. In the simple 
example we are going to consider, the scale factor
is made of two step functions:
\begin{eqnarray}
a^4(\tau)=1+v^2(\theta(\tau+T)-\theta(\tau-T))
\label{salto}
\end{eqnarray}
with $v$ and $T$ being arbitrary parameters. The Bogolyubov coefficients 
provide the following value for the number of created pairs per unit
coordinate volume and unit trimomentum volume in the $k$ mode:
\begin{eqnarray}
<N_k>=\vert \beta_k \vert^2=
\frac{v^4}{4(1+v^2)}\sin^2\left(2Tk\sqrt{1+v^2}\right)
\label{nbogo}
\end{eqnarray}
Expanding the RHS of (\ref{imbo}) using (\ref{nbogo}) up to $\Od(v^4)$ we
find:
\begin{eqnarray}
p_{BOG} \simeq\frac{v^4}{8\pi^2}\int_0^{\infty} dk k^2 \sin^2(2kT)
\label{wbogo}
\end{eqnarray}
The integrand gives the probability density per unit trimomentum volume.
On the other hand, the EA method gives 
the following result from (\ref{frwims}):
\begin{eqnarray}
p_{EA}&\simeq & 2\im \;w 
\simeq 4\frac{1}{32\pi}\frac{1}{72}\int d\tau a^6(\tau)R^2\nonumber \\
&=&
4\frac{v^4}{32\pi}\frac{1}{72}\int d\tau \frac{9}{4}\left(\delta'(\tau-T)
-\delta'(\tau+T)\right)^2 
\end{eqnarray}
We have introduced a global 2 factor in the 
EA because now the field is complex. The spectrum
can be obtained by introducing a complete set of plane waves:
\begin{eqnarray}
p_{EA} &\simeq& 4\frac{1}{32\pi}
\frac{1}{72}\int d\tau d \tau ' \int_{-\infty}^{\infty}
\frac{dp}{2\pi} a^3(\tau)
R(\tau) a^3(\tau ')R(\tau ')e^{-ip(\tau-\tau')}\nonumber \\
&=&\frac{v^4}{64\pi^2}\int_0^{\infty} dp p^2 \sin^2(pT)
\label{imw}
\end{eqnarray}

Comparing $p_{BOG}$ with $p_{EA}$ we find that 
both integrands agree by identifying
$p=2k$. This is sensible and represents the 
energy conservation in  the pair creation,
since $k$ is the single particle energy 
and $p$ is the energy of the gravitational field
oscillations producing particles.

From the complementary point of view, 
given the number density of created particles
$<N_k>$, it is also possible to reconstruct  
the scale factor evolution by inverting the previous
steps: 
\begin{eqnarray}
a^6(\tau)R^2(\tau) &=& 36\left\vert \int_{-\infty}^{\infty}\frac{dp}{2\pi}p
\sqrt{\log(1+<N_{p/2}>)}
e^{-ip\tau}\right\vert^2 +\Od(v^6)\nonumber \\
&=&\frac{9v^4}{4}\left\vert\int_{-\infty}^{\infty}\frac{dp}{2\pi}p
\frac{\sin(pT)}{2} e^{-ip\tau}\right\vert^2 +\Od(v^6)\nonumber \\
&=&\frac{9v^4}{4}\left(\delta'(\tau-T)
-\delta'(\tau+T)\right)^2 +\Od(v^6)
\label{efk}
\end{eqnarray}
This result agrees with the  calculation from (\ref{salto}).

Let us try to generalize the above results for arbitrary 
scale factor evolution.
In the above example, it can be shown that the 
difference in the results using plane waves or a 
complete set of
solutions of the Klein-Gordon equation is $\Od(v^6)$. 
Therefore the former is a good approximation. Now we have to take into account
the presence of the curvature. With that purpose 
we take a variable-frequency plane waves
such that in the vanishing curvature limit they tend to 
the usual plane waves.

Let us consider the temporal part of the Klein-Gordon equation (\ref{tkg}). 
This is a harmonic-oscillator equation but with a time-dependent frequency 
$\omega_k(\tau)=ka^2(\tau)$. Changing to the new 
time coordinate $d\eta=a^2(\tau)d\tau$, the equation can be written as:
\begin{eqnarray}
\frac{d^2\chi_k}{d\eta^2}+2\frac{\dot a}{a}\frac{d\chi_k}{d\eta}+k^2\chi_k=0
\end{eqnarray}
In the limit in which the expansion rate $\dot a/a$ is much smaller than the
 frequency of the oscillations $k$, the equation reduces to the 
 flat space-time form. Therefore let us consider that limit and let us
introduce a complete set of plane waves corresponding to the 
new time coordinate $\eta$:
\begin{eqnarray}
p_{EA} &\simeq& 2 \frac{1}{36}\frac{1}{32\pi}\int d\tau a^6(\tau)R^2(\tau)=
 2 \frac{1}{36}\frac{1}{32\pi}\int d\eta a^4(\eta)R(\eta)^2\nonumber \\
&=& 2 \frac{1}{36}\frac{1}{32\pi}
\int d\eta d\eta' a^2(\eta)a^2(\eta')R(\eta)R(\eta')
\int \frac{dp}{2\pi}e^{-ip(\eta-\eta')}
\label{eaener}
\end{eqnarray}
Changing again to the old coordinate $\tau$ we have:
\begin{eqnarray}
p_{EA} \simeq \frac{1}{8\pi^2}\int_0^\infty dk 
\left \vert\int d\tau\left(\frac{\ddot a}{a^3}
-2\frac{\dot a^2}{a^4}\right)e^{-2ik
\int_0^\tau a^2(\tau')d\tau'} \right\vert^2
\label{eak}
\end{eqnarray}
where we have used (\ref{escrw}). According to the 
above discussion, the introduction of the plane waves only
makes sense in the adiabatic limit and therefore 
this expression is valid only in
the  limit $\dot a/a<<k$. 
As in the step function example, we have used 
$p=2k$. The pairs density $<N_k>$ can be calculated in a very easy way by
identifying (\ref{eak}) with  (\ref{imbo}). Obviously from the equality of
the integrals we cannot obtain the equality of the integrands,
however the
covariance and dimensionality of the integrands allows us to constrain them.
In fact, from the equality of (\ref{eak}) and  (\ref{imbo}) we know that
the integrands differ at most in a function $f(p)$ such that $\int dp f(p)=0$.
Now let us assume that such function can be written as (up to 
$\Od({\cal R}^2)$):
\begin{eqnarray}
f(p)=\int d\eta d\eta' e^{ip(\eta-\eta')} a^2(\eta)a^2(\eta')F(\eta)
\tilde F(\eta')
\end{eqnarray}
where $F(\eta)$ and $\tilde F(\eta')$ are some appropriate functions. 
The condition $\int dp f(p)=0$ implies:
\begin{eqnarray}
\int d\eta a^4(\eta) F(\eta)\tilde F(\eta)=0
\end{eqnarray}
Since the integrand has to be a dimension 4 operator and a scalar function
then the only possibility satisfying that condition is:
\begin{eqnarray}
F\tilde F=\alpha E
\end{eqnarray}
where $E$ is the Gauss-Bonnet term defined before, which is a total
derivative, and $\alpha$ is some arbitrary constant. However we know
that in a radiation dominated universe, where $R=0$, the spectrum is
identically zero $<N_k>=0$ 
(see \cite{Parker1,Parkernew,Parker2}). This implies that the 
contribution from the
Gauss-Bonnet term should also vanish. This  fact allows us to fix the 
constant $\alpha=0$. As a consequence the result in (\ref{eak}) gives the
correct spectrum up to $\Od({\cal R}^2)$ at least in the adiabatic limit
we are considering. In fact, as shown in Table 1 the results are 
in good agreement with
the Bogolyubov
method specially for large values of $k$.

Since we have used an adiabatic approximation in the last
step, we can try to find which are the differences with respect to the 
usual WKB approximation in 
\cite{Brezin}. In this method, the
solutions of the equation (\ref{tkg}) are taken to be:
\begin{eqnarray}
\chi_k(\tau)=\alpha(\tau)e^{-i\psi(\tau)}+\beta(\tau)e^{i\psi(\tau)}
\end{eqnarray}
with
\begin{eqnarray}
\psi(\tau)=k\int_0^\tau d\tau'a^2(\tau')
\end{eqnarray}
with boundary conditons $\alpha(-\infty)=1$, $\beta(\infty)=0$. Putting this
ansatz back into the equation of motion, we get:
\begin{eqnarray}
\dot\alpha e^{-i\psi(\tau)}-\dot \beta e^{i\psi(\tau)}=
-\left(\frac{2\dot a}{a}\right)(\alpha(\tau)e^{-i\psi(\tau)}
-\beta(\tau)e^{i\psi(\tau)})
\end{eqnarray}
where the condition $\dot \alpha e^{-i\psi}+\dot \beta e^{i\psi}=0$ is
used (see \cite{Brezin} for details). 
The solution to first adiabatic order is given by:
\begin{eqnarray}
\beta(\tau)=-\int_\tau^\infty d\tau' \frac{\dot a}{a} e^{-2i\psi(\tau')}
\label{beta}
\end{eqnarray} 
The probability density is given by:
\begin{eqnarray}
p_{WKB}=\frac{1}{2\pi}\int \frac{d^3k}{4\pi^2}\vert \beta(-\infty)\vert ^2
\label{pwbk}
\end{eqnarray}
From (\ref{beta}) and (\ref{pwbk}) we see that 
\begin{eqnarray}
p_{WKB}&=&\frac{1}{8\pi^3}\int d^3k \vert \beta(-\infty)\vert^2
\nonumber \\
&=&
\frac{1}{8\pi^3}\int_0^{\infty} dk 4\pi k^2 \left\vert \int
 d\tau \frac{\dot a}{a}e^{-2ik\int_0^\tau a^2(\tau')d\tau'}\right\vert ^2
\nonumber \\ 
&=&
\frac{1}{2\pi^2}\int dk \left\vert \int  d\tau \frac{\dot a}{a}
k e^{-2ik\int_0^\tau a^2(\tau')d\tau'}\right\vert ^2\nonumber  \\
&=&\frac{1}{2\pi^2}\int dk \left\vert \int  d\tau \frac{\dot a}{a}
\frac{1}{-2ia^2}\frac{d}{d\tau} e^{-2ik\int_0^\tau a^2(\tau')d\tau'}
\right\vert ^2\nonumber \\
&=& \frac{1}{8\pi^2}\int dk \left\vert \int  d\tau \frac{d}{d\tau}\left(
\frac{\dot a}{a^3}\right)
 e^{-2ik\int_0^\tau a^2(\tau')d\tau'}\right\vert ^2
\end{eqnarray}
In the last
step we have used integration by parts assuming that $\dot a/a^3$ vanishes
for $\tau\rightarrow \pm \infty$ (which is the same condition as for
the vanishing of the Gauss-Bonnet term contribution in (\ref{imea})). This condition is 
satisfied in the models
we have considered in the paper. 
Therefore we get:
\begin{eqnarray}
p_{WKB} \simeq \frac{1}{8\pi^2}\int_0^\infty 
dk \left \vert\int d\tau\left(\frac{\ddot a}{a^3}
-3\frac{\dot a^2}{a^4}\right)
e^{-2ik\int_0^\tau a^2(\tau')d\tau'} \right\vert^2
\label{wbk}
\end{eqnarray}
which is valid again only in the adiabatic limit.

\begin{table}
\begin{center}
\begin{tabular}{|c|c|c|c|}
\hline \hline
 k(a.u) & $<N_k>_{BOG}$ & $<N_k>_{EA}$ & $<N_k>_{WKB}$  \\ 
\hline
 $0.3$ & $5.31 \;10^{-4}$& $4.15 \;10^{-4}$ & $4.96 \;10^{-4}$ \\
 $0.4$& $4.31 \;10^{-5}$& $3.33 \;10^{-5}$ & $4.13 \;10^{-5}$ \\
 $0.7$& $2.28 \;10^{-8}$& $1.75 \;10^{-8}$ &$ 2.28 \;10^{-8}$ \\
 $1.0$& $1.22 \;10^{-11}$& $0.93 \;10^{-11}$ & $1.24 \;10^{-11} $\\
 $1.2$& $7.97 \;10^{-14}$& $6.48 \;10^{-14}$ & $8.64 \;10^{-14}$ \\
 $1.3$& $6.44 \;10^{-15}$& $5.91 \;10^{-15}$ & $7.76 \;10^{-15}$ \\
 $1.4$& $4.44 \;10^{-16}$& $4.90 \;10^{-16}$ & $6.19 \;10^{-16}$ \\
\hline \hline
\end{tabular}
\end{center}

\leftskip 1cm
\rightskip 1cm

{\footnotesize {\bf Table 1:} \index{fermions}
Number densities corresponding to the model (\ref{park}) with
 $s=2 (a.u)^{-1}$, $a_1=1$, $a_2=500$ and $b=0$.
BOG denotes the Bogolyubov method and EA the effective action. }

\leftskip 0cm
\rightskip 0cm

\end{table}  
 
In Table 1 some values of the number densities
 are shown for the different methods. The results have been obtained from
 (\ref{eak}) and (\ref{wbk})
for the model (\ref{park})  by numeric integration. 
Due to the strongly oscillating
integrals, the results can only be given for 
small momenta. Both methods give
similar results to those obtained with the 
Bogolyubov coefficients. Notice that, despite the fact that the WKB and EA expression
in (\ref{eak}) and (\ref{wbk}) are not identical, there is no contradiction in this.
These results come from different approximations. The WKB method comes from a derivative
expansion and is not covariant, whereas the EA is an expansion in curvatures and covariance
is imposed from the beginning. From Table 1, we see that the contribution from the 
$(\dot a/a^2)^2$ terms in (\ref{eak}) and (\ref{wbk}) is always smaller than the contribution
from $\ddot a/a^3$. 

\section{Conclusions}

In this work we have shown how to use the 
nonlocal form of the gravitational EA (up to
$\Od({\cal R}^2)$) for the computation
of massless scalar particle production. For FRW backgrounds in which 
the expansion rate asymptotically vanishes, 
it is shown that the particle production probabilities only depend on the
scalar curvature. As a consequence and as expected there is no particle 
creation in a radiation dominated
universe. This is also the case for conformally coupled scalar fields.
For anisotropic homogeneous metrics we reobtain the well-known expression of
Zel'dovich and Starobinski. We compare our results with those 
obtained by means of the well-known Bogolyubov transformations. 
In the examples considered, the
agreement between both methods is complete for the probability densities.
This fact is quite remarkable since we have used a perturbative 
expression for the  EA, whereas the Bogolyubov results are exact.
This indicates that in some cases a perturbative calculation may
contain all the relevant information about the particle 
production processes. In principle, the EA is defined for
asymptotically flat manifolds, however it is interesting to notice
that the naive extension to those manifolds in which adiabatic
vacua can be defined, properly reproduces the correct results.
Finally we have also compared the different spectra 
derived with the EA, Bogolyubov 
and WKB techniques.

In principle the EA method can be extended to more general metrics 
(not necessarily 
homogeneous) in a straightforward way. This fact could make it 
valuable in
those areas in which the Bogolyubov technique has been traditionally used. 
In addition this method can also be applied to the production of higher 
spin particles such as Dirac and Weyl fermions, gravitons, gravitinos, etc. 
Finally, in a recent work \cite{Vilko2} the relevance of  
the nonlocal EA for 
particle creation 
has also been stressed from a different point of view based on the
energy-momentum tensor expectation values.

\vspace{0.5cm}
{\bf Acknowledgements:} A.L.M acknowledges support from SEUID-Royal Society.
This work has been partially supported by Ministerio de
Educaci\'on y Ciencia (Spain) CICYT (AEN96-1634).

\end{document}